\documentclass[aps,prl,showpacs,twocolumn,amsmath,amssymb]{revtex4-1}

\usepackage{graphicx}% Include figure files
\usepackage{dcolumn}% Align table columns on decimal point
\usepackage{bm}% bold math
\usepackage{amsmath}
\usepackage{psfrag}
\usepackage{subfigure}

\newcommand{\sign}{\mathrm{sign}}
\newcommand{\Eq}[1]{Eq.~(\ref{#1})}
\newcommand{\Eqs}[1]{Eqs.~(\ref{#1})}
\newcommand{\la}{\langle}
\newcommand{\ra}{\rangle}
\begin{document}

%\preprint{preprint}

\title{Reentrance effect in macroscopic quantum tunneling \\and non-adiabatic
Josephson dynamics in $d$-wave  junctions}%

\author{J. Michelsen}

\author{V.S. Shumeiko}
\affiliation{%
Department of Microtechnology and Nanoscience, MC2, Chalmers University of Technology,
SE-41296 Gothenburg, Sweden
}%

%\date{6 May 2010}%


\begin{abstract}

We develop a theoretical description of non-adiabatic Josephson dynamics in superconducting
junctions containing low energy quasiparticles. Within this approach we investigate the effects of midgap states in junctions of unconventional d-wave superconductors. We identify a reentrance effect in the transition between thermal activation and macroscopic quantum tunneling, and connect this phenomenon to the experimental observations in Phys. Rev. Lett. 94, 087003 (2005).  It is also shown that nonlinear Josephson dynamics can be defined by resonant interaction with midgap states reminiscent to nonlinear optical phenomena in media of two-level atoms.
\end{abstract}

\maketitle

%%%%%%%%%%%%%%  Introduction %%%%%%%%%%%%%%%

With the advent of superconducting qubits
\cite{Makhlin2001,Wendin2007,Clark2008} a general
interest has grown towards realization of macroscopic
quantum dynamics in superconducting
weak links. The superconducting qubits developed
so far are based on Josephson tunnel junctions of conventional superconductors. 
A conceptually interesting and practically important question is whether other types of
Josephson weak links, such as junctions of high temperature superconductors, and mesoscopic 
metallic or semiconducting weak links can be employed in qubit circuits.
The central aspect of this problem is to understand the role of low energy electronic states 
usually present in such junctions. The low energy quasiparticles are driven away from
equilibrium by temporal variation of the superconducting phase across the junction,
and produce a non-adiabatic contribution to the Josephson
current. This effect is commonly considered to result
in dissipation, and decoherence of qubit states. However,
examples from nonlinear optics show that resonant interaction
with {\em localized} electronic states (two-level atoms)
may generate a nonlinear dispersion of electromagnetic
modes leading to a variety of nonlinear phenomena involving coherent energy exchange 
between macroscopic and microscopic variables \cite{Eberly}.
This kind of nonlinear phenomena, whose origin differs
from the nonlinearity of the adiabatic Josephson
potential, has never been studied in the context
of macroscopic Josephson dynamics.

In this Letter we investigate the non-adiabatic Josephson
dynamics in artificial grain boundary junctions of
high temperature superconductors \cite{Mannhart2002}, which is caused
by interaction with superconducting surface bound states
(midgap states). The midgap states (MGS) situate at
zero energy in the middle of the superconducting energy gap \cite{Hu1994},
and are fundamentally connected to the unconventional d-wave superconducting
order parameter in these materials \cite{VanHarlingen1995,Tsuei2000}. We find
that interaction with the MGS has implications in both the imaginary time
dynamics (tunneling) and the real time nonlinear dynamics of the
junction. First, we show that the MGS are capable of 
significantly affecting the transition between the thermal activation and
macroscopic quantum tunneling (MQT) decay of Josephson current state inducing
multiple, forward and backward, transitions between the two regimes. We suggest 
that such a reentrance phenomenon underlines the experimentally observed 
\cite{Bauch2005} anomaly of the switching current rates.
Secondly, we show that the nonlinear resonant response of
d-wave junctions may be entirely caused by the nonlinear dynamics of the MGS,
and lead to a bifurcation regime with an explosive growth of the response
amplitude. These findings are made within the framework of a general theoretical
description of the non-adiabatic Josephson dynamics in junctions containing low
energy quasiparticles, developed in this paper.

The special role of the MGS is explained by their discrete energy spectrum,
and pairwise coupling to the temporal variation of the superconducting phase.
Tunneling spectroscopy data \cite{Covington1997} as well as observation of a
$\pi$-junction transition \cite{Testa2005} provide experimental evidence for
the MGS existence. The equilibrium properties of MGS and their role in the dc
Josephson effect are well studied in the literature (see reviews
\cite{Kashiwaya2000,Lofwander2001} and references therein). The multiple
degenerate zero energy level of the MGS splits into a narrow band under the
effects of tunneling and anisotropy of the d-wave order parameter,
$\Delta({\mathbf k_F}) =\Delta_0\cos(2\theta)$. Due to the small bandwidth a 
thermal saturation of the MGS occurs at relatively low temperatures that may be 
comparable to the MQT transition temperature. This saturation
effect accompanied by the decrease of the MGS-induced dissipation underlines,
as we show, the reentrance effect in the MQT transition. In junctions  with
atomically smooth interfaces, a large fraction of tunneling electron
trajectories contains hybridized MGS pairs. The two-state Rabi dynamics  and
the MGS saturation at large driving amplitudes define the nonlinear property
of real time Josephson dynamics.

%%%%%%%%%%%%%%%% Instability equation  %%%%%%%%%%%%%%%%%%%%

{\em  MQT transition temperature.}  %
 We start with the discussion of the effect of MGS on the
MQT transition temperature. We follow the method of Ref. \cite{Grabert1984},
based on the analysis of the imaginary time dynamics of phase fluctuations, $\delta\varphi(\tau)$, around the steady phase difference across the junction,  $\varphi=\varphi_b$, at the top of the barrier of the tilted Josephson potential. In this method, the MQT transition is
manifested by an instability of the phase fluctuations described with an
effective euclidian action, $S_\text{eff}[\varphi]\approx
S_\text{eff}[\varphi_b]+ \sum_n
\Lambda(\varphi_b,i\nu_n)\delta\varphi_n\delta\varphi_{-n}$, $\nu_n = 2\pi
nT$ ($k_B=\hbar = 1$). The transition corresponds to the change of the sign
of the kernel, $\Lambda(\varphi_b,i\nu_1)$, and the temperature is given by
the equation $\Lambda(\varphi_b, i\nu_1) = 0$.

To derive the effective action for the superconducting phase, we consider the
partition function of d-wave junction, $Z=\int
\mathcal{D}\varphi\mathcal{D}^2\psi
\,e^{-(S_\varphi[\varphi]+S_\psi[\varphi,\psi])}$, and perform integration
over fermionic variables $\psi$ \cite{Ambegaokar1982}. Here $S_\varphi =\int
d\tau [(C/8e^2)\dot{\varphi}^2 - I_\text{e}\varphi/2e]$ is the macroscopic
part of the action contributed by the charging energy of the junction capacitance, $C$, and    
the inductive energy of the biasing current, $I_e$. Furthermore, $S_\psi =\int d\tau \int dr    
\ \bar{\psi}(\partial_\tau + \mathcal{H} + (i/4)\sign(x)\dot{\varphi})\psi$  is the
microscopic part of the action, associated with the mean-field Hamiltonian of
the superconducting electrons, $\mathcal{H} $, the last term provides electro-neutrality within the electrodes \cite{ZAZ2}.

We perform the integration by choosing a general method suitable for all
kinds of junctions regardless of their transparencies or presence of
localized surface states. We separate the spatial problem from the temporal
one by introducing a basis of {\em instantaneous} eigenstates of electronic
Hamiltonian, $\mathcal{H}\phi_i= E_i\phi_i$,
$\psi(\mathbf{r},\tau)=\sum_i\phi_i(\mathbf{r};\varphi) a_i(\tau)$.
The Fermionic action then becomes, $S_\psi=\int d\tau \sum_{ij}
\bar{a}_iG^{-1}_{ij}a_j$, where $G^{-1}_{ij}=\partial_\tau+ {H}_{ij}
(\varphi,\dot{\varphi})$, is the inverse Green's  function of the effective
Hamiltonian,
\begin{eqnarray}\label{Hij}
{H}_{ij} &=& E_i\delta_{ij}-i\dot{\varphi}\mathcal{A}_{ij};\\
\label{Aij} {A}_{ij} &=& \left(\phi_i,i\partial_\varphi\phi_j
\right)-(1/4)(\phi_i,\sign(x)\sigma_z\phi_j)
\end{eqnarray}
is the matrix element of quasiparticle transitions induced by temporal
variation of the phase. The effective action has the form,
$S_\text{eff}[\varphi]= S_\varphi - \text{Sp}\ln\hat{G}^{-1}$.

The saddle point solution is given by equation, $\delta S_\text{eff} =0$. For
the fermionic contribution we have, $\delta\text{Sp}\ln G^{-1}
=(1/2e)\text{Sp}\left(\hat{I}_J\hat{G}\delta\varphi\right)$, where
\vspace{-0.25cm}
\begin{equation}\label{Ij}
\hat{I}_J=2e\left(\partial_\varphi
\hat{E}+i[\hat{E},\hat{\mathcal{A}}]\right)
\vspace{-0.2cm}
\end{equation}
is the Josephson current operator \cite{ZAZ2}, see Appendix. At the static
saddle point, $-\hat{G}^0(\tau,\tau)= \hat n^0(\hat E)$ is the equilibrium
density matrix commuting with $\hat E$, therefore only the diagonal
(adiabatic) part of the current operator contributes to the Josephson
current, $I_J^{ad}(\varphi)= 2e\sum_i\partial_\varphi E_i n^0_i$, that
defines $\varphi_b$, $I_J^{ad}(\varphi_b) - I_{\text{e}}=0$.

The non-adiabatic effect is described by the second functional derivative of
the fermionic action, $(1/2e)^2 \text{Sp}\left(\delta\varphi\hat{I}_J
 \hat{G}^0\hat{I}_J\hat{G}^0\delta\varphi\right)$, and the fluctuation kernel
reads (see Appendix), $\Lambda(i\nu_n) =
(C/8e^2)\left(\nu_n^2-\omega_b^2-i\nu_n \gamma_0(i\nu_n)\right)$. Here
$\omega_b^2=-(2e/C)\partial_{\varphi} I_J^{ad}$ is the plasma frequency at
the barrier, and
\vspace{-0.2cm}
\begin{equation}\label{gamma0}
\gamma_{0}( i\nu_n)=\frac{4e^2}{C}\sum_{ij}
\frac{\varepsilon_{ij}|\mathcal{A}_{ij}|^2(n_i^0-n_j^0)}{\varepsilon_{ij}-i\nu_n},
\vspace{-0.2cm}
\end{equation}
is the quasiparticle response; $\varepsilon_{ij}=E_i-E_j$, $n_i^0= n_F(E_i)$
is the Fermi filling factor, all functions are taken at $\varphi= \varphi_b$.

Up to this point the derivation is general, and \Eq{gamma0} applies to all
the quasiparticles. At small frequencies, however, only the MGS and itinerant
nodal quasiparticles \cite{Scalapino1995} are  relevant. Furthermore, the MGS contribution has more pronounced temperature dependence compared to the nodal states because MGS have a small bandwidth, $\varepsilon_m \ll \Delta_0$.  
%added line on nodal QP vs MGS
Focusing on the more interesting effect of the MGS,
 we truncate \Eq{gamma0} to the MGS subspace. The matrix
elements, $\mathcal{A}_{ij}$, only connect MGS pairs of the same
electronic trajectory while transitions among the trajectories are forbidden
due to preserved translational invariance. Parameterizing the MGS pairs with 
%added "more" info on \theta 
the angle, $\theta$, between the incidental wave vector $\mathbf{k}_F$ of the respective trajectory and the interface normal (see top inset Fig 1), and denoting,
$\varepsilon(\theta) = E_1(\theta) -E_2(\theta)$,
$A(\theta)=\mathcal{A}_{12}$, we present the equation for the transition
temperature on the form,
\begin{equation}\label{Tk}
\nu^2- \omega_b^2-{8e^2S\over  C} \nu^2 \left\langle{\varepsilon
A^2(n_1^0-n_2^0)\over\varepsilon^2+\nu^2 } \right\rangle = 0,
\end{equation}
where angle brackets indicate the average over the Fermi surface, $S$ is the
junction area, $\nu= 2\pi T$.

The temperature dispersion of the MGS term in \Eq{Tk} is primarily defined by
the Fermi filling factors and the resonant denominator, while the particular form
of the smooth functions $\varepsilon(\theta)$ and $A(\theta)$ plays a secondary
role. This allows us to formulate an {\em analytical} model equation for the
transition temperature, thus circumventing the difficulty of evaluating
anisotropy of the MGS, which generally can only be done numerically. By
replacing $\varepsilon A^2(\theta)(d\varepsilon/d\theta)^{-1}$ with a constant, we get
\Eq{Tk} on the form, $F(x) = \varepsilon_m^2x^2(1+\eta f(x))- \omega_b^2 =
0$, where $f(x) = \int_0^1 dy \tanh(\pi y/2x)(x^2+y^2)^{-1}$, and
$x=\nu/\varepsilon_m$; $\eta=8a\pi/R_nC\varepsilon_m$ is the coupling
strength, $R_n = \pi/e^2S\langle D\rangle$ is the normal junction resistance,
and $a \sim 1$ is a geometry specific constant. The latter estimate is
obtained from the scaling, $\varepsilon_m \propto \sqrt D\Delta_0$, and $A\propto
\sqrt D$, in the limit of small transparency,  $D\ll 1$, extracted from the
analytical equations for the MGS spectrum and transition matrix elements
, see Appendix. The advantageous property of this analytical model is that it
applies to junctions with interface faceting, which is taken into account by
average values of the model parameters, $\eta$, $\varepsilon_m$, and
$\omega_b$.

\begin{figure}[h!]
\begin{center}
\includegraphics{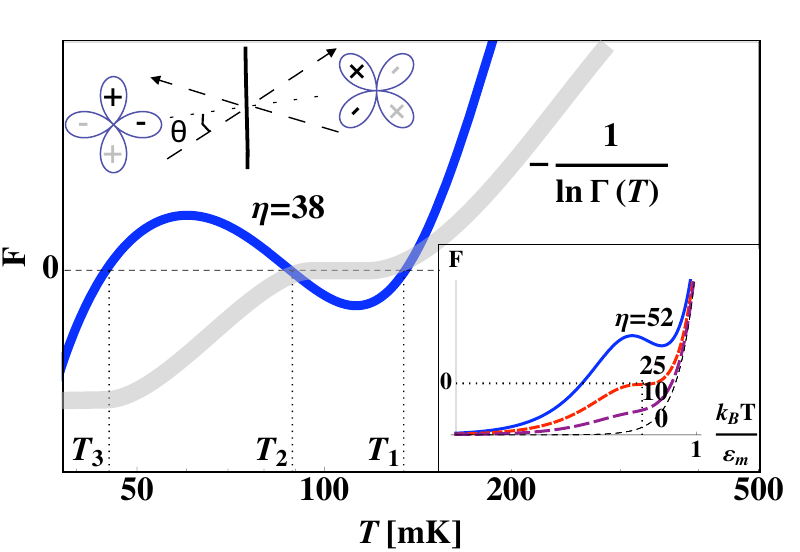}\vspace{-0.3cm}
\caption{Reentrance effect in MQT. Sketch of temperature dependence of decay
rate (wide shadow line) illustrates the effect featuring three transitions
between thermal activation and MQT regimes. Experimental transition
temperatures are given by zeros of function $F(x)$, defined in the text (blue line) for $\eta = 38$. Lower inset shows development of non-monotonic feature of function $F(x)$ with increasing $\eta$, at $\eta >25$. Upper inset illustrates junction geometry and scattering electron trajectory (dashed line). \vspace{-0.8cm}
}\label{reentrance}
\end{center}
\end{figure}

Numerical solutions to the modeled \Eq{Tk} are presented in the inset to Fig.
\ref{reentrance}. They demonstrate splitting of a single critical point
into three critical points manifesting the reentrance effect. The bifurcation of the solution
to \Eq{Tk} occurs at the coupling strength, $\eta=25$, and the barrier frequency, $\omega_b= 3.45\,\varepsilon_m$. This phenomenon can be understood as a reentrance effect: At high temperature the thermal activation undergoes a transition to MQT in the absence of interaction with MGS since the MGS are saturated; with lowering temperature, the MQT rate decreases because of increased interaction with MGS, and thermal activation takes over; then it undergoes the second transition to MQT in the presence of interaction. This finding constitutes the first main results of this paper.

In the experiment with a tilt YBCO junction \cite{Bauch2005} an anomalous temperature dependence of the Josephson current decay rate has been observed, which can be interpreted
in terms of the reentrance effect: transition to the MQT regime at $T_1\approx 135$ mK is interrupted, at
$T_2\approx 90$ mK, by reentrance of the thermal activation, which then
undergoes the second MQT transition at $T_3\approx 45$ mK, as sketched on
Fig. \ref{reentrance}. To make a quantitative comparison we fit the three
experimental transition temperatures by adjusting the average model parameter
values, $\eta$, $\varepsilon_m$, and $\omega_b$, see Appendix, as shown
on Fig. \ref{reentrance}. Including the stray $LC$-oscillator of the
experimental setup \cite{Bauch2006} does not make any qualitative difference
but rather insignificantly (within 20\%) shifts the parameters values. The
best fit is eventually achieved for the values, 
%$\eta\approx 45$,
$\varepsilon_m\approx 320$ mK, $\omega_b\approx 1.7$ K, $\omega_p \approx 2.5$ K, and $C\approx 36$fF, assuming experimental values of the critical current, $I_C=1.4\mu$A, and the switching current, $I_e\approx 0.9I_C$. Given the experimental junction transparency, $D\sim 10^{-4}$, we are able to evaluate the maximum energy gap at the interface, $\Delta_0\approx 16$ K. The
geometrical constant in the equation for $\eta$ is estimated for the
experimental value $R_n = 500\,\Omega$,  $a\approx 1.5$, as expected. 

In our discussion the temperature dependence of the {\em adiabatic} Josephson potential has been ignored. This dependence, also originating from the thermal saturation of the MGS band, may play a role in junctions with large capacitance 
where it may modify, as shown in  \cite{Bauch2005}, the thermally activated decay rate and provide an alternative explanation to the experimentally observed feature.
%Comp with experiment (to be revised).

Consistency of our {\em non-adiabatic} reentrance scenario with the experimental observations strongly indicates involvement of the MGS pairs in the macroscopic dynamics of the junction. Moreover, it provides us with valuable information about the microscopic MGS parameters. 

%\vspace{-0.4cm}
%

%\vspace{-0.4cm}

%%%%%%%%%%%%%%%%%%%%%%%%%%%%%%%%%%%%%%%% Non-linear resonance %%%%%%%%%%%%%%%%%
%
{\em Nonlinear resonance Josephson dynamics.} %
To investigate the real time Josephson dynamics, one needs to generalize our
approach to non-equilibrium states. This is done by considering the partition
function defined through the action on the real time Keldysh contour
\cite{Zaikin1990}. Then proceeding, as before, by introducing the
instantaneous basis, we derive the equation for the Keldysh-Green's functions,
$\hat G^{ab}$,
%
%\begin{equation}\label{GFdef}
$[i\partial_t-\hat{H}(\varphi^a,\dot\varphi^a)]
\hat{G}^{ab}(t-t')=a\delta^{ab}\delta(t-t')$,
%\end{equation}
%
with the same Hamiltonian as in \Eq{Hij},  here $a,b=\pm$ label the forward
and backward branches of the Keldysh contour. The semiclassical dynamics of
the superconducting phase is given by the least action principle,
$(\delta/\delta \chi) S_\text{eff}[\varphi,\chi]_{\chi=0} = 0$, formulated in
terms of the Wigner variables, $\varphi^a=\varphi+a \chi/ 2$
\cite{Kamenev2005}. Calculating
 the functional derivative, we get,
\begin{equation}\label{eqphi}
{C\over 2e}\,\ddot\varphi + \text{Tr}\left(\hat{I}_J\hat{\rho} \right)
={I_e}, \quad \hat{I}_J  = 2e(\partial_{\varphi} \hat{E} +
i[\hat{E},\hat{\mathcal{A}}]).
\end{equation}
Here $\hat{\rho}(t)=(1/2i)\sum_a  \hat{ G}^{aa}(t,t)$ is the non-equilibrium
single particle density matrix, which satisfies, by virtue of the equation
for $\hat G^{ab}$, the Liouville equation,
\begin{equation}\label{Liouville}
i \dot{\hat{\rho}}=[\hat H\,,\,\hat{\rho}], \quad \hat H = \hat{E}
-\dot{\varphi}\hat{\mathcal{A}}.												%changed A to caligraphic A
\end{equation}
Eqs. (\ref{eqphi}) and (\ref{Liouville}) are exact in the semiclassical
limit, and give a general description of the non-adiabatic Josephson dynamics
in all kinds of junctions. These equations constitute another main result of
this paper.

%%%%%%%%%%%%%%%%%%%%%%%%%%%%%%%%%%%%%
For the MGS pairs, \Eq{Liouville} reduces to the  Bloch equation for the
two-level density matrix parameterized with the angle $\theta$. In this case,
\Eqs{eqphi}, (\ref{Liouville}) become analogous to the ones describing
electromagnetic modes in a cavity embedded in a medium of two-level atoms
\cite{Eberly}. The most interesting is the case of the resonant excitation of
the MGS pairs, which corresponds to the Josephson plasma frequency lying
within the MGS band, $\omega_p <\varepsilon_m$. Suppose a small oscillating
biasing current is applied to the junction, $I_e\cos\omega t$, with frequency
slightly detuned from the plasma frequency, $\delta = \omega - \omega_p \ll
\omega$. The resonant dynamics of the  superconducting phase,
$\varphi(t)=\text{Re}(\varphi_\omega e^{-i\omega t})$, is described by the
averaged equation for slow varying complex amplitude, $\varphi_\omega(t)$,
\begin{equation}\label{slowphi}
-2i\dot{\varphi}_\omega +\left[-2\delta
+\gamma(r)\right]\varphi_\omega=eI_e/\omega_p C,
\end{equation}
where $\gamma=\gamma' +i\gamma''$ is the nonlinear MGS response,
\begin{equation}\label{gamma}
\gamma'(r)= \gamma_0'
+\partial^2_{\varphi}\bar{\varepsilon}{\gamma''\over\Gamma_1}\, r^2, \;
\gamma''(r) \!=\! \frac{\Gamma\gamma_0''}{\sqrt{(r\bar{A}\omega)^2+\Gamma^2}}
\end{equation}
(the nonlinear adiabatic term is dropped from \Eq{slowphi} to emphasize the
MGS effect). In \Eq{gamma} the bar indicates the										%moved def of r and bar to earlier
resonant values, $r=|\varphi_\omega|$, and the quantity $\gamma_0$ refers to the
linear MGS response given by the analytical continuation of \Eq{gamma0} to
 real frequencies, $i\nu\rightarrow \omega+i0$. The response is calculated (see Appendix) by solving the Bloch equation (\ref{Liouville}) assuming the MGS 
adiabatically following, in the rotating frame, the evolution of the phase amplitude, 
and adding small decoherence rates $\Gamma_1, \Gamma_2\ll \varepsilon_m$. 
The MGS decoherence is induced, e.g., by scattering to the itinerant nodal states by 
the facet edges or other rare inhomogeneities, leading to the MGS intrinsic broadening,
$\Gamma =\sqrt{\Gamma_1\Gamma_2}$. The dissipative part of the linear response 
is estimated,
\vspace{-0.4cm}
\begin{equation}\label{gamma0res}
\gamma_0''(\omega, T)  \sim {\omega\over \varepsilon_m R_nC}\tanh{\omega\over
4T}.
%\quad \omega_p <\varepsilon_m,
\end{equation}
It gives the frequency independent quality factor at zero temperature,
$Q_{MGS}= \omega/\gamma_0'' \sim \varepsilon_m R_nC$. It is instructive to
compare this result to the damping effect of the nodal quasiparticles,
$Q_{nodal} \sim \Delta_0 R_nC \gg Q_{MGS}$, see Appendix (cf.
\cite{Bruder1995,Barash1995,Kawabata2005}).

\begin{figure}[h!]
\begin{center}
\vspace{-0.4cm}
\includegraphics{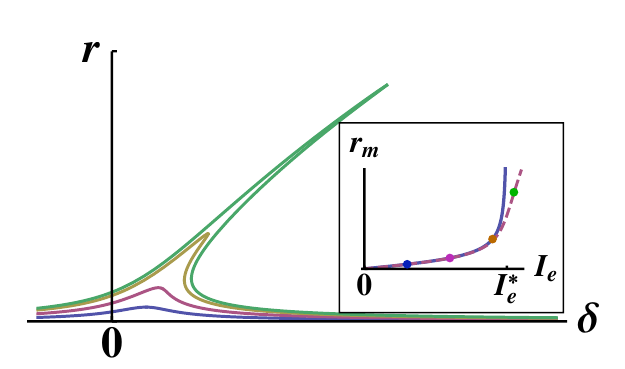}\vspace{-0.4cm}
\caption{Effect of MGS on nonlinear resonance response of the junction.
Response amplitude as function of detuning is shown for different amplitudes
of driving current.  Inset: maximum response amplitude as a function of
driving current, dots indicate current values in the main figure.\vspace{-0.8cm}
}\label{solutions}
\end{center}
\end{figure}

Equation (\ref{gamma}) provides an extension of the linear response equation
(\ref{gamma0}) to the nonlinear region, when the Rabi frequency of MGS
transitions exceeds the MGS intrinsic width, $r\bar A\omega \gtrsim \Gamma$.
In this nonlinear regime relevant for narrow MGS levels  the stationary
response amplitude as function of detuning is defined by relation,
\begin{equation}
\delta=\frac{1}{2}\gamma'(r)\pm
\frac{1}{2r}\sqrt{(e{I}_e/C\omega_p)^2-[\gamma''(r) r]^2}.
\end{equation}
The response demonstrates the bifurcation regime shown in Fig.
\ref{solutions}, which is typical for nonlinear oscillators, but here is
entirely controlled by MGS characteristics rather than adiabatic Josephson
potential. The bifurcation appears at very small driving currents, $\tilde
I_e = (I_e/2I_C) Q_{MGS}\sim\Gamma/\omega\bar A\ll 1$. The most striking
feature of the response is the explosive growth of the peak amplitude,
%
%\begin{equation}\label{rmax}
$r_{max}= \tilde I_e [ 1- ( \tilde I_e /
I_e^\ast)^2]^{-1/2}$,
%\end{equation}
%
for the driving current approaching the value $ I_e^\ast=\Gamma/\omega\bar
A$. This effect is caused by the MGS saturation at large driving amplitudes,
which is manifested by decreasing damping in equation (\ref{gamma}). The
divergency is smeared by adding small damping, e.g., by nodal quasiparticles,
and changes to a steep dependence asymptotically approaching the line,
$r_{max}= (\tilde I_e- I_e^\ast)(Q_{nod}/Q_{MGS})$. The Rabi dynamics of the
MGS should be more clearly exposed in the time resolved experiments.

%\vspace {-0.4cm}
%
%
%\vspace{-0.5cm}

%%%%%%%%%%%%%%%%%%%%%%%%%%%%%%%%%%%%%%%%%
In conclusion, we considered the effects of midgap states on Josephson
dynamics in d-wave superconducting junctions. The analysis is based on the
developed general theoretical framework for non-adiabatic Josephson dynamics
in junctions containing low energy quasiparticles.  We identified a
reentrance effect in MQT, and connected that to the experimental
observations. We also investigated the nonlinear dynamical response of the
junction caused by coupling to nonlinear MGS dynamics.
By analyzing the experiment \cite{Bauch2005} in terms of the interaction with
MGS, we found that the MGS bandwidth in the experimental junction is smaller
than the Josephson plasma frequency, $\varepsilon_m<\omega_p$. This implies
that the resonance condition for MGS excitation is not fulfilled, and MGS
should not  affect the real time junction dynamics, which thus would be
similar to conventional Josephson oscillators. The quality factor is then
defined by the nodal quasiparticles, and is estimated from the experimental
data, $Q_{nod}  \sim (\Delta_0/\omega_p)^2 \sim 40$. In order to increase
this factor the strategy would be to increase the ratio, $\Delta_0/\omega_p$,
which is, however, impossible beyond the limit, $Q_{nod}\sim 1/D$, provided
MGS remain off-resonance ($\varepsilon_m\sim \sqrt D \Delta_0 <\omega_p$).
Exceeding this limit necessarily implies resonant excitation of the MGS,
and establishing the nonlinear regime described in this paper.

%\vspace{-0.05cm}
{\em Acknowledgement.} We are thankful to J. Clark, M.
Fogelstr\"om, T. L\"ofwander, and C. Tsuei for useful discussions;
illuminative discussion of experiment with Th. Bauch and F. Lombardi are
gratefully acknowledged. The work was supported by the Swedish Research
Council (VR), and the European FP7-ICT Project MIDAS.

\newpage

\appendix

%% supplementary material 6-JM ,
%% Mar 8 2010 JM
%
%
%\documentclass[aps,prb,amsmath,widetext,amssymb]{revtex4}
%%\documentclass[aps,prl,showpacs,amsmath,twocolumn,amssymb]{revtex4}
%%\documentclass[onecolumn,showpacs,preprintnumbers,amsmath,amssymb]{revtex4}
%%\documentclass[preprint,showpacs,preprintnumbers,amsmath,amssymb]{revtex4}
%
%
%
%\usepackage{graphicx}% Include figure files
%\usepackage{dcolumn}% Align table columns on decimal point
%\usepackage{bm}% bold math
%\usepackage{amsmath}
%\usepackage{psfrag}
%\usepackage{subfigure}
%
%\newcommand{\bra}[1]{\langle #1 |}
%\newcommand{\ket}[1]{| #1 \rangle}
%\newcommand{\braket}[2]{\langle #1 | #2\rangle}
%\newcommand{\sign}{\mathrm{sign}}
%\newcommand{\Eq}[1]{Eq.~(\ref{#1})}
%\newcommand{\Eqs}[1]{Eqs.~(\ref{#1})}
%\newcommand{\la}{\langle}
%\newcommand{\ra}{\rangle}
%%\nofiles
%
%\begin{document}

%\preprint{preprint}
\widetext{
\section{Appendix}%: \\Reentrance effect in macroscopic quantum tunneling and non-adiabatic Josephson dynamics in $d$-wave junctions.
%}%
%
%\author{J. Michelsen}
%%\email{jens.michelsen@mc2.chalmers.se}
%\author{V.S. Shumeiko}
%\affiliation{%
%Department of Microtechnology and Nanoscience, MC2, Chalmers University of
%Technology, SE-41296 Gothenburg, Sweden
%}%
%
%%\date{6 March 2010}%
%
%\begin{abstract}
%{
In this appendix we present details of the derivation of (a) effective action for superconducting phase, multiple critical
temperatures for transitions between thermal activation and MQT regimes under
interaction with MGS; (b) MGS energy spectrum 
and transition matrix elements, MGS
linear response and comparison to the damping effect of nodal quasiparticles; and (c) the
nonlinear junction dynamics under resonant interaction with MGS. 
%There are few notational changes introduced here compared to the
%article: we shall use the convention $\hbar=k_B=1$, and the transition matrix elements
%will be denoted by a cursive $\mathcal{A}$.
%}
%\end{abstract}
%%\pacs{74.50.+r, 74.45.+c, 71.70.Ej}%
%\maketitle
%
%  Reentrance Effect in MQT
%%%%%%%%%%%%%%%%%%%%%%%%%%%%%%%%%%%%%%%%%%%%%
%
\section{Reentrance effect in MQT}
In this section we derive the dispersion equation for small phase fluctuation
in imaginary time used to evaluate the crossover temperature between thermal
activation and MQT decay of persistent Josephson current. To this end
 we shall also need to derive equations defining the MGS characteristics: energy dispersion equation, 
 and interlevel matrix elements, and discuss MGS general properties, and present some explicit analytical
equations.

\subsection{Imaginary time approach}
Starting from the imaginary time representation of partition function and performing
integration over the fermionic fields, one obtains the equilibrium partition function as
a path integral over the phase with an euclidian effective action $S_E^{\text{eff}}$,
\begin{equation}
Z=\int \mathcal{D}\varphi e^{-S_E^\text{eff}[\varphi]}, \qquad
S_E^\text{eff}[\varphi]=\int_0^\beta d\tau\left[
\frac{C}{8e^2}\dot{\varphi}^2+U_\text{ext}(\varphi)\right]-\text{Sp}\ln\hat{G}^{-1},
\end{equation}
where $U_\text{ext}(\varphi)=-(I_\text{e}/2e)\varphi$ is an inductive energy
of a biasing current $I_e$, and
\begin{equation}
\hat{G}^{-1}_{ij}=\Bigl(\partial_\tau+\hat{H}_{ij}(\varphi,\dot{\varphi})\Bigr), \qquad \hat{H}_{ij}=\delta_{ij}E_i-i\dot{\varphi}\mathcal{A}_{ij}.
\end{equation}
The matrix $\hat E$ in this equation is constructed with the eigen
energies, $E_{ij}=E_i\delta_{ij}$ of the microscopic junction Hamiltonian,
\begin{equation}\label{HE}
\mathcal{H}\phi_i = E_i\phi_i,
\end{equation}
\begin{equation}\label{BdGhamiltonian}
\mathcal{H}= \left[\frac{(-i\mathbf{\nabla})^2}{2m}-\mu+V\right]\sigma_z+
\hat{\Delta}e^{i\chi}\sigma_++\hat{\Delta}e^{-i\chi}\sigma_-,
\end{equation}
where $\hat{\Delta}$ denotes a non-local operator, $\hat{\Delta}\phi\equiv\int
d\mathbf{r}' \Delta(\mathbf{r},\mathbf{r}')\phi(\mathbf{r}') =\int
d\mathbf{k}\,\Delta(\mathbf{k},\mathbf{r})\int
d\mathbf{R}\,\phi(\mathbf{r}+\mathbf{R})e^{i\mathbf{k}\cdot\mathbf{R}}$, and
\begin{equation}
\chi(\mathbf{r})\equiv \begin{cases}
\frac{\varphi}{2}, \ &\mathbf{r}\in L\\
-\frac{\varphi}{2}, \ &\mathbf{r}\in R
\end{cases},
\end{equation}
denotes the phase of the order parameter in the left ($L$) and right ($R$) electrodes;
$V(\mathbf{r})$ represents an interface potential.

The matrix of operator $\mathcal{\hat A}$ is constructed with the matrix elements of the
transition between the basis states, \Eq{HE},
\begin{equation}
\mathcal{A}_{ij}=(\phi_i, \left[i\partial_\varphi-(1/4)\sigma_z\sign(x)\right]\phi_j).
\end{equation}
An alternative representation of this matrix is given through the relation to the matrix
$I_{ij}=(\phi_i,\,\hat I_J\,\phi_j)$,
\begin{equation}\label{transmat}
i\mathcal{A}_{ij}=\frac{1}{2e}\frac{I_{ij}}{\varepsilon_{ij}}, \qquad i\neq j,
\end{equation}
where $\varepsilon_{ij}=E_i-E_j$. As it is shown in the next section, this matrix
represents the Josephson current flowing through the junction, and is connected to the
current density operator via equations,
\[I_{ij}=\int_S d\mathbf{n}\cdot \mathbf{j}_{ij}(\mathbf{r}),\]
\begin{equation}\label{j}
\mathbf{j}_{ij}(\mathbf{r})=\frac{e}{2m}\left.
(i\mathbf{\nabla}-i\mathbf{\nabla}')\phi_i^\dag(\mathbf{r})
\phi_j(\mathbf{r}')\right|_{\mathbf{r}=\mathbf{r'}}.
\end{equation}

\subsection{Current operator}\label{currentop}
Here we present a proof for the interpretation of $\hat{I}_J$ as the quantum mechanical
current operator. To do so we identify the current density operator as
\begin{equation}\label{j}
\mathbf{j}_{ij}(\mathbf{r})=\frac{e}{2m}\left.(i\mathbf{\nabla}-i\mathbf{\nabla}')\phi_i^\dag(\mathbf{r})\phi_j(\mathbf{r}')\right|_{\mathbf{r}=\mathbf{r'}}.
\end{equation}
The current through the interface, $S$, is given by \[I_{ij}=\int_S d\mathbf{n}\cdot
\mathbf{j}_{ij}(\mathbf{r}).\] If we consider the system as an infinitely large loop we
can use the fact that no current is flowing through any other part of the surface of the
superconductor so we may extend the surface $S$ around the whole superconductor and use
Gauss law:
\begin{equation}\label{Ij}
2I_{ij}=\int_{\mathbf{r}\in L} dV \  \mathbf{\nabla}\cdot
\mathbf{j}_{ij}(\mathbf{r})-\int_{\mathbf{r}\in R} dV \  \mathbf{\nabla}\cdot
\mathbf{j}_{ij}(\mathbf{r}).
\end{equation}
From the explicit form of the BdG Hamiltonian (\ref{BdGhamiltonian}) one finds the
relations,
\begin{equation}\begin{split}
(-i\mathbf{\nabla})\cdot\mathbf{j}_{ij}(\mathbf{r})&=-\frac{e}{2m}\left[[-\nabla^2\phi_i(\mathbf{r})]^\dag\phi_j(\mathbf{r})-\phi_i^\dag(\mathbf{r})[-\nabla^2\phi_j(\mathbf{r})]\right]\\
   &=-e\left[(E_i-E_j)\phi_i^\dag(\mathbf{r})\sigma_z\phi_j(\mathbf{r})+\phi^\dag_i(\mathbf{r})[\mathcal{H},\sigma_z]\phi_j(\mathbf{r})\right].\\
\end{split}
\end{equation}
The last term (the commutator between the "charge operator" $\sigma_z$, and the
quasiparticle Hamiltonian) can be rewritten as
\begin{equation}
[\sigma_z,\mathcal{H}]=2 i\partial_{\chi} \mathcal{H}=\begin{cases}
4i\partial_\varphi \mathcal{H},  \ & \mathbf{r}\in L\\
-4i\partial_\varphi \mathcal{H},  \ & \mathbf{r}\in R\\
\end{cases}.
\end{equation}
The current operator then becomes
\begin{equation}
I_{ij}=(-i)2e\left[(\phi_i,i\partial_\varphi\mathcal{H}\phi_j)-(E_j-E_i)\frac{1}{4}(\phi_i,\sign(x)\sigma_z\phi_j)\right].
\end{equation}
By differentiating the eigenvalue equation $\mathcal{H}\phi_i=E_i\phi_i$ wrt $\varphi$
one obtains the following identities:
\begin{equation}\begin{split}
&(\phi_i,i\partial_\varphi\mathcal{H}\phi_i)=i\partial_\varphi E_i,\\
&(\phi_i,i\partial_\varphi\mathcal{H}\phi_j)=(E_j-E_i)(\phi_i,i\partial_\varphi\phi_j),
\quad i\neq j
\end{split}.
\end{equation}
From this one sees that the current matrix elements are given by
\begin{equation}\begin{split}
I_{ii}&=2e\partial_\varphi E_i,\\
I_{ij}&=2e i(E_i-E_j)\mathcal{A}_{ij},
\end{split}
\end{equation}
or in explicit matrix form
\begin{equation}
\hat{I}=2e\left(\partial_\varphi\hat{E}+i[\hat{E},\hat{A}]\right).
\end{equation}
%

%%%%%%%%%%%%%%%%%%%%%%%%%%%%%%%%%%%%%%%%%%%%%%%%%%%%%%%%%%%
\subsection{Quasiclassical wave functions}\label{quasiclassical}
%
%%%%%%%%%%%%%%%%%%%%%% HTS6 %%%%%%%%%%%%%%%%%%%%%%%%%%%%%%%%%%%%%

In this subsection we sketch the quasiclassical formalism used for the evaluation of the
MGS properties.

Consider an interface with normal, $\hat{\mathbf{n}}$, pointing in the positive $x$
direction ($\hat{\mathbf{n}}\cdot\hat{\mathbf{x}}>0$). The incident angle, $\theta$, of
an electronic  trajectory is defined through the relation
\[k_F\cos\theta=\mathbf{k}_F\cdot \mathbf{n}.\]
The d-wave order parameter is aligned with the crystal a-b axes according to
$\Delta_0(k_a^2-k_b^2)$ where $k_a=\hat{\mathbf{k}}_F\cdot \hat{\mathbf{a}}$,
$k_b=\hat{\mathbf{k}}_F\cdot \hat{\mathbf{b}}$. Introducing the misorientation angle
$\alpha:$ $\mathbf{n}\cdot\mathbf{a}=\cos\alpha$ and
$\mathbf{n}\cdot\mathbf{b}=\sin\alpha$ we can write $k_a=\cos(\alpha-\theta)$ and
$k_b=\sin(\alpha-\theta)$. The order parameter can then be written as a function of the
two angles $\alpha,\theta$:
\begin{equation}
\Delta(\theta)=\Delta_0\cos2(\alpha-\theta).
\end{equation}
Assuming specular reflection, the momentum parallel to the interface is conserved upon
reflection, while the perpendicular momentum is inverted,  (or equivalently, the
reflected angle is given by $\pi-\theta$). The quasiclassical wave functions have the
form of linear combinations of plane waves
\begin{equation}
\phi(\mathbf{r},\mathbf{k}_{F||})=\frac{1}{\sqrt{S}}e^{i\mathbf{k}_{F||}
\cdot\mathbf{r}_{||}}\sum_{\sigma=\pm}\tilde{\phi}^\sigma(x)
e^{i\mathbf{k}_F^\sigma\cdot\mathbf{n} x},
\end{equation}
where $\mathbf{r}=x\mathbf{n}+\mathbf{r}_{||}$, $\mathbf{k}_F^\sigma=\sigma
\mathbf{k}_{F\perp}+\mathbf{k}_{F||}$, and $S$ is the interface area. The slowly varying
envelopes, $\tilde{\phi}^\sigma(x)$, satisfy the quasiclassical BdG equations,
\begin{equation}\label{quasiclass}
\left[\mathbf{v}^\sigma_F\cdot\mathbf{n}(-i\partial_{x})\sigma_z+\Delta_{j}^\sigma
e^{i\chi}\sigma_++\Delta_{j}^\sigma
e^{-i\chi}\sigma_-\right]\tilde{\phi}^\sigma(x)=E\tilde{\phi}^\sigma(x),
\quad j=\begin{cases}
L, \quad x<0\\
R, \quad x>0
\end{cases},
\end{equation}
where the shorthand notation is introduced,
$
\Delta_j^\sigma=%\begin{cases}
\Delta_0\cos 2\left(\sigma\theta-\alpha_j\right),% \quad j=L\\
%\Delta_0\cos 2\left(\sigma\theta-\alpha_R\right), \quad j=R
%\end{cases}
%\end{equation}
$
and $\alpha_{j=L,R}$ denote the angles of the a-b crystal axes to the normal of the
interface, $\hbar=1$. It is convenient to incorporate the sign of the order parameter
into the phase,
\begin{equation}
\varphi_j^\sigma=\begin{cases}
\varphi/2+\Theta(-\Delta_L^\sigma)\pi, \quad &j=L\\
-\varphi/2-\Theta(-\Delta_R^\sigma)\pi, \quad &j=R
\end{cases}.
\end{equation}
The local interface potential is replaced in the quasiclassical approximation with the
boundary conditions for slow wave functions envelops,
\begin{equation}\label{boundarycond}
\phi^\sigma(0^-)=T^{\sigma\sigma'}\phi^{\sigma'}(0^+),
\end{equation}
where $T^{\sigma\sigma'}$ denotes a general single channel (i.e. for given trajectory)
transfer matrix, characterized by the transmission amplitude, $d(\theta)$, and
reflection amplitude $r(\theta)$, ($|r|^2+|d|^2=1$).
%where $s_j^\sigma=\sign[\Delta_j^\sigma]$.
%
%%%%%%%%%%%%%%%%%%%%%%%%

%
\subsection{MGS spectrum and transition matrix elements}\label{mgsspectrum}
Here we derive equations defining the MGS energy spectrum and transition matrix elements
for planar junctions with specular interfaces.

The bound state solutions to equation (\ref{quasiclass}) have the general form,
\begin{equation}\label{boundstatewf}
\tilde{\phi}^\sigma(x)=\begin{cases}
A^\sigma u_L^\sigma e^{-\zeta_L^\sigma|\tilde{x}|}, \quad x<0\\
B^\sigma u_R^\sigma e^{-\zeta_R^\sigma|\tilde{x}|}, \quad x>0
\end{cases},
\end{equation}
where
\begin{equation}
u^\sigma_j=\frac{1}{\sqrt{2}}\begin{pmatrix}
e^{-i\gamma_j^\sigma}\\
e^{i\gamma_j^\sigma}
\end{pmatrix}, \quad
\zeta_j^\sigma=\frac{|\Delta_j^\sigma|}{|\mathbf{v}_F^\sigma
\cdot\mathbf{n}|}\sin\eta_j^\sigma, \quad j=L,R,
\end{equation}
and $\gamma_j^\sigma=\sigma\eta_j^\sigma+\varphi_j^\sigma/2$ with
\begin{equation}
2\eta_j^\sigma=\begin{cases}
\text{acos}(E/|\Delta_l^\sigma|), \quad j=L\\
-\text{acos}(E/|\Delta_r^\sigma|), \quad j=R
\end{cases}.
\end{equation}
Using the properties of the spinors, the matching condition Eq. (\ref{boundarycond}) can be
rewritten into two sets of equations
$A^\sigma=\mathcal{N}^{\sigma\sigma'}B^{\sigma'}$,
$0=\mathcal{M}^{\sigma\sigma'}B^{\sigma'}$
where
$ \mathcal{N}^{\sigma\sigma'}=
T^{\sigma\sigma'}\cos\left(\gamma_L^\sigma-\gamma_R^{\sigma'}\right), \
\mathcal{M}^{\sigma\sigma'}=
T^{\sigma\sigma'}i\sin\left(\gamma_L^\sigma-\gamma_R^{\sigma'}\right) $,
determining the coefficients, $A^\sigma$, and, $B^\sigma$, upto a normalization constant.
The condition that these equations have non-trivial solutions,
$\text{det}\mathcal{M}=0$, defines the spectral equation,
\begin{equation}
\prod_{\sigma=\pm} \sin\left[\gamma_L^\sigma(E)-\gamma_R^\sigma(E)\right]=
R\prod_{\sigma=\pm}\sin\left[\gamma_L^\sigma(E)-\gamma_R^{-\sigma}(E)\right].
\end{equation}
where $R=|r|^2=1-D$.

To obtain an expression for the transition matrix elements, $\mathcal{A}_{12}$, we make
use of the  relationship with the current matrix elements in Eq. (\ref{transmat}),
\[\mathcal{A}\equiv i\mathcal{A}_{12}=\frac{1}{2e}\frac{I_{12}}{(E_1-E_2)},\]
where  the current matrix elements are given by \Eq{j} within the quasiclassical
approximation,
\begin{equation}\label{I12}
I_{12}=e\sum_{\sigma=\pm}\left.(\mathbf{v}_F^\sigma\cdot\mathbf{n})
(\tilde{\phi}^\sigma_1)^\dag\tilde{\phi}_2^\sigma\right|_{x=0}= e\sum_{\sigma=\pm}
(\mathbf{v}_F^\sigma\cdot\mathbf{n}) [B^\sigma_1]^*B^\sigma_2
\cos\left[\eta^\sigma_R(E_1)-\eta^\sigma_R(E_2)\right].
\end{equation}
%

%%%%%%%%%%%%%%%%%%%%%%%%%%%%%%%%%%%

\subsection{Selection rule}
For a junction with $\frac{\pi}{4}/\frac{\pi}{4}$ orientation, there exists a symmetry
relation, $\Delta_L^\sigma=\Delta_R^\sigma\equiv\Delta_{\frac{\pi}{4}}$, and
consequently, $\eta_L^\sigma=-\eta_R^\sigma\equiv
\eta_\frac{\pi}{4}=(1/2)\text{acos}(E/\Delta_\frac{\pi}{4})$. The spectral equation then
simplifies,
\begin{equation}
\cos^2(2\eta_\frac{\pi}{4})=D\cos^2(\varphi/2),
\end{equation}
and has the two solutions $E_1=-E_2=\sqrt{D}\Delta_\frac{\pi}{4}\cos(\varphi/2)$. For a
spatially symmetric potential, the amplitudes in \Eq{boundstatewf} are  given by
equations (upto normalization factor ensuring that $ (\phi,\phi)=1$),
\begin{equation}\label{45amplitudes}\begin{aligned}
B^+_{1,2}&=\sqrt{R}\cos(\varphi/2),  \quad   &A^+_{1,2}&=\pm \sqrt{R}\cos(\varphi/2)\\
B^-_{1,2}&=\sin(2\eta_\frac{\pi}{4}(E_{1,2})+\varphi/2), \quad &A^-_{1,2}&=\pm \sin(2\eta_\frac{\pi}{4}(E_{1,2})+\varphi/2)
\end{aligned}
\end{equation}
Inserting these amplitudes into equation (\ref{I12}) for the current matrix element
(which is proportional to $\mathcal{A}$), we arrive at the important result,
$$\mathcal{A}=0.$$

Now we show that this result is a particular case of a general selection rule forbidding
transitions among the MGS for {\em any symmetric junction}. This selection rule is
imposed by the symmetry of the Hamiltonian, Eq. (\ref{BdGhamiltonian}), under charge and
parity conjugation $\mathcal{CP}$,
\[\mathcal{CP}\phi(\mathbf{r})=\phi^*(-\mathbf{r}).\]
To prove our statement we first note that the $\mathcal{C}\mathcal{P}$-symmetry splits
the Hilbert space of the Hamiltonian (\ref{BdGhamiltonian}), into two subspaces which
correspond to the even and odd transformations of the eigen states under
$\mathcal{C}\mathcal{P}$ conjugation,
\[\mathcal{CP}\phi_i=\pm \phi_i.\]
Then we find that the operator that defines the  transition matrix elements,
$\mathcal{A}=i\mathcal{A}_{12}=
-(\phi_1,[\partial_\varphi+(i/4)\text{sign}(x)\sigma_z]\phi_2)$,  respects the
$\mathcal{C}\mathcal{P}$-symmetry, and therefore the matrix elements between the states
belonging to the odd-subspace and even-subspace vanish.

Next, we notice that an arbitrary symmetric junction is obtained by continuous rotation of
the ${\pi\over 4}/{\pi\over 4}$ junction. Such a rotation preserves the
$\mathcal{C}\mathcal{P}$-symmetry, and the eigen functions transform smoothly under the
rotation, unless the nodes of the order parameter are crossed. Therefore,  the wave
functions initially belonging to different discrete subspaces of the symmetry
operator will maintain this property during the rotation. Inspection of Eqs.
(\ref{boundstatewf}) and (\ref{45amplitudes}) for the ${\pi\over 4}/{\pi\over 4}$
junction proves that indeed the two MGS eigen functions obtain opposite signs under
$\mathcal{C}\mathcal{P}$ transformation, and thus belong to different subspaces of the
symmetry operator.

This proves that the transition matrix elements will equal zero for the MGS of all symmetric
junctions.

\subsection{$\frac{\pi}{4}+\kappa /\frac{\pi}{4}-\kappa$ orientations}
The antisymmetric orientation is one of the few orientations  for which one can obtain
non-trivial analytical solutions $\mathcal{A}$. For these orientations the symmetry
holds,
\begin{equation}
|\Delta_L^\pm|=|\Delta_R^\mp|\equiv \Delta^{\pm}.
\end{equation}
For trajectories that admit a pair of MGS we find the spectral equation,
\begin{equation}
\cos^2(\eta^++\eta^-)=D\cos^2\varphi/2.
\end{equation}
where the shorthand $\eta^\pm=\eta_L^\pm=-\eta_R^\mp$ was
introduced for notational convenience.
Using the definitions, $\cos2\eta^\pm=E/|\Delta^\pm|$, and
$\sin2\eta^\pm=\sqrt{1-E^2/|\Delta^\pm|^2}$, we
find the solution,
\begin{equation}\label{eq.4545antisym}
E=\pm|\Delta^+\Delta^-|\sqrt{D}\cos\frac{\varphi}{2}
\frac{2\sqrt{1-D\cos^2\frac{\varphi}{2}}}{\sqrt{(|\Delta^+|+
|\Delta^-|)^2-4|\Delta^+\Delta^-|D\cos^2\frac{\varphi}{2}}}.
\end{equation}
Once an analytical expression for the spectrum has been found one can
also obtain an analytic expression for $\mathcal{A}$ in terms of $\eta^\pm$,
\begin{equation}
\mathcal{A}_{12}=\sqrt{R}
\frac{(\sin2\eta^+-\sin2\eta^-)}{2E\sin\left(\eta^++\eta^-\right)}
\left(\frac{|\Delta^+\Delta^-|\sin2\eta^+\sin2\eta^-}{|\Delta^+|\sin2\eta^+
+|\Delta^-|\sin2\eta^-}\right).
\end{equation}
Here $\eta^\pm=(1/2)\arccos(E/|\Delta^\pm|)$ with the energy given by Eq.
(\ref{eq.4545antisym}).

Notice that for $\kappa=0$,  we have $\eta^+=\eta^-$, so that the matrix element
vanishes for this orientation, as also shown above.  For small misorientation,
$\kappa\ll 1$, this expression can be expanded into
\begin{equation}
\mathcal{A}_{12}\approx
\frac{\sqrt{RD}\cos\frac{\varphi}{2}}{\sqrt{1-D\cos^2\frac{\varphi}{2}}} \frac{\delta
\Delta}{\Delta},
\end{equation}
where $\delta \Delta(\theta)= \Delta^+(\theta)-\Delta_0\sin2\theta\approx 2\kappa
\Delta_0\cos 2\theta$. This equation reduces at small transparency,
\begin{equation}
\mathcal{A}_{12}\approx \sqrt{D}\cos\frac{\varphi}{2} \frac{\delta
\Delta}{\Delta}\approx 2\kappa \sqrt{D}\cot(2\theta).
\end{equation}

%
%
%
%%%%%%%%%%%%%%%%%%%%%%%%%%%%%%%%%%%%%%%
%
%

\subsection{Transition temperature }
The decay of the persistent Josephson current at large bias current applied to the
junction is represented by escape of a fictitious particle representing the junction
from a metastable potential well of the "washboard potential" formed by the periodic
Josephson potential, $U_J(\varphi)$, and potential of the current bias,
$U_\text{ext}(\varphi)=-(I_e/2e)\varphi$. Grabert and Weiss \cite{Grabert1984} devised a
method for direct calculation of a critical temperature of transition from the thermally
activated escape to the escape via MQT by analyzing small fluctuations around the saddle
point located at the barrier top, $\varphi_b$. The semiclassical euclidian action is
expanded to second order in the deviation, $\delta\varphi=\varphi-\varphi_b$,
\begin{equation}\label{schmidt}\begin{split}
S_E^\text{eff}[\varphi] &\approx S^{(0)}[\varphi_b]+S_{E}^{(2)}[\delta\varphi],\\
S_{E}^{(2)}&=\int_0^\beta d\tau\int_0^\beta d\tau' \delta\varphi(\tau)\Lambda(\tau-\tau')\delta\varphi(\tau')\\
&=\sum_n \Lambda(i\omega_n)|\varphi_n|^2.
\end{split}
\end{equation}
The Fourier components, $\Lambda(i\nu_n)$, of the fluctuation kernel,
$\Lambda(\tau-\tau')$, with Matsubara frequencies, $\nu_n=2\pi n/\beta$, are then the
eigenvalues associated with the gaussian fluctuations around the stationary point,
$\varphi_b$. In the thermal activation regime, the stationary point is stable, and all
the eigenvalues are positive, $\Lambda(i\omega_n)>0$. Transition to the MQT regime is
manifested by the instability, indicated by the sign change of the smallest eigenvalue,
$\Lambda(i\nu_1)< 0$. The transition temperatures can thus be obtained by finding
solutions to the equation, $\Lambda(i\nu_1)=0$.

To evaluate the fluctuation part of the action for our system we expand in
$\delta\varphi$ and keep only second order terms. This gives us,
\begin{equation}
S_{E}^{(2)}[\delta\varphi]=\int_0^\beta d\tau
\left[\frac{C}{8e^2}\delta\dot{\varphi}^2+\frac{1}{2}\left. \frac{\partial^2
U_\text{ext}}{\partial\varphi^2}\right|_{\varphi_b}\delta\varphi^2\right]
+\int^{\beta}_{0} d\tau d\tau' \delta\varphi(\tau)K(\tau-\tau')\delta\varphi(\tau'),
\end{equation}
where
\begin{equation}
K(\tau-\tau')=\frac{1}{2}\left.\frac{\delta^2\text{Sp}\ln
\hat{G}^{-1}}{\delta\varphi(\tau)\delta\varphi(\tau')}\right|_{\varphi=\varphi_b}.
\end{equation}
It is convenient to perform the functional
differentiation in a basis where the dependence on $\dot{\varphi}$ is removed. This is
achieved by a using rotation matrix $U: \partial_\varphi U=i\mathcal{A}U$,
\begin{equation}
K(\tau-\tau')=\frac{1}{2}\text{Tr}
\left(\tilde{\hat{\rho}}(\tau)\frac{\partial^2\tilde{\hat{H}}}{\partial\varphi^2}
(\tau)\right)_{\varphi_b}\delta(\tau-\tau')+\text{Tr}\left(\tilde{\hat{G}}(\tau,\tau')
\frac{\partial\tilde{\hat{H}}}{\partial\varphi}(\tau')\tilde{\hat{G}}(\tau',\tau)
\frac{\partial\tilde{\hat{H}}}{\partial\varphi}(\tau)\right)_{\!\varphi_b}.
\end{equation}
The contribution from the first part combines with the second derivative of the external
potential to define the barrier frequency,
\begin{equation}
-\omega_b^2=\frac{8e^2}{C}\left[\frac{1}{2}\frac{\partial^2 U_\text{ext}}{\partial\varphi^2}+\frac{1}{2}\text{Tr}\left(\tilde{\hat{\rho}}(\tau)\frac{\partial^2\tilde{\hat{H}}}{\partial\varphi^2}(\tau)\right)\right]_{\varphi_b},
\end{equation}
leaving the second part which we denote by small $k(\tau-\tau')$:
\begin{equation}
k(\tau-\tau')=\frac{1}{(2e)^2}\text{Tr}\left(\hat{G}(\tau,\tau')\hat{I}_J(\tau')\hat{G}(\tau',\tau)\hat{I}_J(\tau)\right)_{\varphi_b}.
\end{equation}
Here $\hat{I}_J$ is the current operator as defined in Eq. (\ref{phaseeqm}), the
imaginary time (Matsubara) Green function for constant phase ($\varphi(\tau)=\varphi_b$)
is given by equation,
\begin{equation}
(G_0)_{ij}(\tau,\tau')=-\left[\theta(\tau-\tau')(1-\rho_i^0)-
\theta(\tau'-\tau)\rho_i^0\right]e^{-E_i^0(\tau-\tau')}\delta_{ij},
\end{equation}
where $\rho_i^0=n_F(E_i^0)$, and $E_i^0=E_i(\varphi_b)$. Due to the boundary condition,
$\delta\varphi(0)=\delta\varphi(\beta)$, we can perform integration by parts to obtain,
\begin{equation}
S_{E}^{(2)}[\delta\varphi]=\int_0^\beta d\tau d\tau' \delta\varphi(\tau)\Lambda(\tau-\tau')\delta\varphi(\tau'),
\end{equation}
where
\begin{equation}
\Lambda(\tau-\tau')=\frac{C}{8e^2}\left[\left(-\partial_\tau^2 -\omega_{b}^2\right)\delta(\tau-\tau')+\frac{8e^2}{C}k(\tau-\tau')\right].
\end{equation}
The Fourier representation of the operator $\Lambda (\tau-\tau')$ defines the
eigenvalues,
\begin{equation}
\Lambda(i\nu_n)=\frac{C}{8e^2}\left[-(i\nu_n)^2-
\omega_{b}^2-i\nu_n \gamma_{0}(i\nu_n)\right], \qquad \nu_n=\frac{2\pi n}{\beta},
\end{equation}
where $-i\nu_n\gamma_0(i\nu_n)=(8e^2/C)k(i\nu_n)$ has the explicit form,
\begin{equation}\label{gamma0Im}
\gamma_{0}(i\nu_n)=\frac{4e^2}{C}\sum_{ij}
\frac{\varepsilon_{ij}|\mathcal{A}_{ij}|^2(\rho_i^0-\rho_j^0)}{\varepsilon_{ij}
-i\nu_n}.
\end{equation}
Here $\varepsilon_{ij}=E_i^0-E_j^0$ and $\mathcal{A}_{ij}=\mathcal{A}_{ij}(\varphi_b)$.
Comparing this result with Eq. (\ref{schmidt}) we find the equation for the transition
temperature to the MQT regime,
\begin{equation}\label{equationT}
\Lambda(i\nu_1)\propto \nu_1^2-\omega_b^2-i\nu_1 \gamma_0(i\nu_1)=0.
\end{equation}

\subsection{MGS and reentrance effect}\label{crossover-reentrance}
The reentrance effect described in the article results from a strong temperature
dependence of dissipation produced by the MGS, which decrease with increasing
temperature. After truncating to the MGS subspace, we present
\Eqs{gamma0Im} and (\ref{equationT}) on the form, dropping the subscript,
\begin{equation}\begin{split}
&\nu^2-\omega_b^2- i\nu\gamma_0(i\nu_1)=0\\
& \Rightarrow \nu^2-\nu\frac{4e^2S}{C}\sum_{\pm}\left\langle \frac{i\varepsilon\mathcal{A}^2\rho_z}{\pm\varepsilon-i\nu}\right\rangle=\omega_b^2\\
& \Rightarrow \nu^2\left(1+ \frac{8e^2S}{C}\left\langle \frac{\varepsilon\mathcal{A}^2\rho_z}{\varepsilon^2+\nu^2}\right\rangle\right)=\omega_b^2.\\
%& \Rightarrow T^2\left(1+\frac{8e^2}{C}\left\langle \frac{\varepsilon\mathcal{A}^2\rho_z(T)}{\varepsilon^2+(2\pi T)^2}\right\rangle\right)=T_0^2
\end{split}
\end{equation}
Here $S$ is the junction area, and $\rho_z=n_F(E_1)-n_F(E_2)$ while the average $\langle\ldots\rangle$ is defined as
\begin{equation}
\langle\ldots \rangle=\int_+ \frac{d^2\mathbf{k}_F}{(2\pi)^2}\ldots,
\end{equation}
where integration is performed over the Fermi wave vectors in
the positive direction of the interface normal.
Assuming the interface to be orthogonal to the crystal a-b plane, and taking into
account strong anisotropy of the Fermi surface, we write the integral on the form,
\begin{equation}
S\langle \ldots \rangle=N\int_{-\pi/2}^{\pi/2} d\theta \ldots\ldots,
\end{equation}
where $N=Sk_F/2\pi c$ is the number of conducting channels for a stack 2D planes with
spatial period $c$.
%
%\noindent \textbf{Simple Model:}
%
For the sake of simplicity, we consider almost symmetric MGS spectrum,
$E_1=-E_2=\varepsilon/2$, giving $\rho_z(T)=\tanh(\varepsilon/4T)$, and proceed to
integration over $\varepsilon$ in the integral over $\theta$,
The equation for the crossover temperature then becomes,
\begin{equation}
(2\pi T)^2\left(1+\frac{8e^2}{C}N\int_0^{\varepsilon_m}  d\varepsilon g(\varepsilon)
\frac{\varepsilon \mathcal{A}^2(\varepsilon)\tanh(\varepsilon/4T)}{\varepsilon^2+ (2\pi
T)^2}\right)=\omega_b^2.
\end{equation}
where $g(\varepsilon)=4d\theta/d\varepsilon$ is the MGS spectral density.
The important qualitative features of the integral, independent of junction geometry,
are the saturation effect due to the MGS population number, $\tanh(\varepsilon/4T)$, and
the resonance feature in the denominator, $\varepsilon^2+(2\pi T)^2$. Numerical studies
show that these features define the temperature dependence of the integral, while the
role of the function, $g(\varepsilon)\varepsilon\mathcal{A}^2(\varepsilon)$, which
contains information about junction geometry, is qualitatively insignificant. This
observation allows us to approximate the latter with some constant whose magnitude is set by $\mathcal{A}^2\sim D$, because $g(\varepsilon)\sim 1/\varepsilon_m$, and $\varepsilon\sim\varepsilon_m$;
\begin{equation}
g(\varepsilon)\varepsilon\mathcal{A}^2(\varepsilon)= a \int_{-\pi/2}^{\pi/2} d\theta D(\theta),
\end{equation}
where $a$ is a geometry dependent numerical constant of order $\sim 1$. We are then able to formulate
a simple model equation defining the transition temperature,
\begin{equation}\label{bigdispersioneq}
(2\pi T)^2\left(1+\frac{8\pi a}{R_nC} \int_0^{\varepsilon_m}
d\varepsilon\frac{\tanh(\varepsilon/4T)}{\varepsilon^2+(2\pi T)^2}\right)=\omega_b^2,
\end{equation}
where $R_n=\pi/e^2S\la D\ra$ is the normal junction resistance.

\subsection{Fitting MQT transition temperatures }
Here we shall outline the method used to fit the transition temperatures in our model to the experiment in \cite{Bauch2006}. For this procedure, Eq. (\ref{bigdispersioneq}) will be our model. Before proceeding we first simplify our notation in Eq. (\ref{bigdispersioneq}) by writing $\gamma_0(i\nu)=i\nu\eta f(\nu/\varepsilon_m)$, with $f(x) = \int_0^1 dy \tanh(\pi y/2x)(x^2+y^2)^{-1}$, where
$\eta=8a\pi/R_nC\varepsilon_m$ is the coupling strength.
The next, crucial step is to consider a dimensionless function, $F(x,\eta)
=x^2(1+\eta f(x))$, where $x=2\pi T/\varepsilon_m$, and
choose the scaling parameter $\varepsilon_m$ such that the three argument
values, corresponding to given transition temperatures, $T_1,\,T_2$, and
$T_3$ give the same function value, $F(x_1,\eta)=F(x_2,\eta)=F(x_3,\eta)$;
this can only be  achieved by adjusting simultaneously the shape of the
function $F(x,\eta)$ by tuning parameter $\eta$. This procedure gives unique
values for both parameters. Then the barrier frequency is determined by
equating, $\omega_b^2=\varepsilon_m^2F(x_1,\eta)$. 

It should be noted that in our analysis we have neglected the temperature dependence of the adiabatic Josephson potential. In general the saturation of the MGS may lead to strong temperature dependence of the Josephson current - a feature suggested in \cite{Bauch2005} to be the origin of the hump structure. The model was that the potential barrier height changes between two asymptotically temperature independent values over a narrow region 100 mK $<T<$ 150 mK, assumed to still be in the thermally activated regime. The temperature dispersion of the decay rate corresponding to the two different barrier heights is indicated in their Fig. 2b by two shifted lines. This explanation was consistent with an MQT crossover temperature $T^*=50$ mK obtained from the plasma frequency with estimated junction capacitance $C=1$ pF. Later experiments\cite{Bauch2006}, however, suggested that this value of the junction capacitance was overestimated due to the presence of a stray capacitance originating from the STO substrate. Comparison with typical grain boundary junctions would suggest a junction capacitance of the order of 100 fF, thus increasing the crossover temperature to values right around the anomalous features of the temperature dispersion of the decay rate. Therefore, while the mechanism suggested in \cite{Bauch2005} could produce a feature like the one observed in their experiments, the parameters of this particular junction suggest that the reentrance effect discussed in the present paper may be more relevant.

In addition to the stray capacitance from the substrate it was argued\cite{Bauch2006} that the c-axis transport in the tilted junction may cause a stray inductance. We can include the effect of such a stray $LC$ oscillator in our analysis by adding an extra term, $\lambda x^2/(x^2 + \tilde\omega_0^2)$, to the function $F(x,\eta)$, where
$\tilde\omega_0= \hbar\omega_0/\varepsilon_m$ is the dimensionless frequency
of the stray oscillator, and $\lambda = \hbar^2/L_0C\varepsilon_m^2$ is the
coupling containing the stray inductance and (unknown) capacitance $C$ of the
junction. The latter is connected to the barrier frequency through the
relations, $\omega_b = \omega_p(1-(I_e/I_C)^2)^{1/4}$, and
$\omega_p=2eI_C/\hbar C$, and evaluated through an iteration procedure,
assuming switching current $I_e\approx 0.9I_C = 1.26 \,\mu$A. Including the
$LC$ oscillator does not produce any qualitative changes but rather slightly
modifies numerical values of the fitting parameters.
The
best fit is eventually achieved for the values, 
%$\eta\approx 45$,
$\varepsilon_m\approx 320$ mK, $\omega_b\approx 1.7$ K, $\omega_p \approx 2.5$ K, and $C\approx 36$fF, assuming experimental values of the critical current, $I_C=1.4\mu$A, and the switching current, $I_e\approx 0.9I_C$. Given the experimental junction transparency, $D\sim 10^{-4}$, we are able to evaluate the maximum energy gap at the interface, $\Delta_0\approx 16$ K. The
geometrical constant in the equation for $\eta$ is estimated for the
experimental value $R_n = 500\,\Omega$,  $a\approx 1.5$, as expected. 

%
%
%
%%%%%%%%%%%%%%%%%%%%%%%%%%%%%%%%%%%%%%
%
%

\section{Linear response}
In this section we investigate the different processes contributing to the linear damping in d-wave Josephson junctions.
We start by presenting the expression for the linear response in a general form, and
then evaluate the contribution to the linear dissipation coming from the MGS to MGS
transitions and compare that with the contributions coming from competing processes of
nodal to nodal state transitions, and  MGS to nodal state transitions.

The non-adiabatic, real-time dynamics in Josephson junctions  is described by the dynamical
equations governing the evolution of the superconducting phase,
\begin{equation}\label{phaseeqm}
\frac{C}{2e}\ddot{\varphi}+I_J^\text{ad}(\varphi)+
\text{Tr}\left(\hat{I}_J(\hat{\rho}-\hat{\rho}^0)\right)=I_e(t), \qquad
\hat{I}_J=2e\left(\partial_\varphi
\hat{E}+i\left[\hat{E},\hat{\mathcal{A}}\right]\right),
\end{equation}
and the single quasiparticle density matrix,
\begin{equation}\label{densitymatrixeqm}
i\partial_t\hat{\rho}=[\hat{H},\hat{\rho}], \qquad
\hat{H}=\hat{E}-\dot{\varphi}\hat{\mathcal{A}}.
\end{equation}
In Eq. (\ref{phaseeqm}), we have subtracted the adiabatic component of the Josephson
current, $I_J^\text{ad}(\varphi)=\text{Tr}(\hat{I}_J\hat{\rho}^0)$, where $\hat{\rho}^0$
denotes the initial density matrix, and added an external current, $I_e(t)$ of the
biasing circuit. 

The effects discussed in the article concern small oscillations around a stationary
point $\varphi_0$.  Straightforward linearization of Eqs. (\ref{phaseeqm}) and
(\ref{densitymatrixeqm}) with respect to small deviations from the equilibrium,
$\varphi-\varphi_0$, $\rho-\rho^0$ lead to the dispersion equation,
\begin{equation}
\left(-\omega^2+\omega_p^2+\omega\gamma_0(\omega)\right)\varphi_\omega=0,
\end{equation}
where $\gamma_0(\omega)$ denotes the linear response of the quasiparticles,
\begin{equation}\label{linearresponse}
\gamma_0(\omega)=\frac{4e^2}{C}\sum_{ij}
\frac{\mathcal{A}_{ij}^2\varepsilon_{ij}(\rho_{ii}^0-\rho_{jj}^0)}{\varepsilon_{ij}-
(\omega+i0)},
\end{equation}
The indices $i,j,$ here refer to continuous and discrete sets of quantum numbers
characterizing the different eigen states, $\phi_i$, of the microscopic Hamiltonian.

%%%%%%%%%%%%%%%%%%%%%%%%%%%%%%%%%%%%%%%%%
\subsection{MGS to MGS transitions}\label{mgslinresp}

Here we shall make use of the general expression for the linear response, Eq.
(\ref{linearresponse}), to evaluate the MGS contribution  for nearly symmetric junctions
$\kappa\ll 1$. The contribution due to MGS transitions has the form,
\begin{equation}\label{mgslinear}
\gamma_0(\omega)=\frac{4e^2N}{C}\sum_\pm\int_{-\pi/2}^{\pi/2}d\theta\frac{\varepsilon
\mathcal{A}^2\rho_z^0}{\pm\varepsilon-\omega-i0}.
\end{equation}
%
%\begin{equation}\label{mgslinear}
%\gamma_0(\omega)=\frac{4e^2}{C}\sum_\pm\left\langle
%\frac{\varepsilon\mathcal{A}^2\rho_z^0}{\pm\varepsilon-\omega-i0} \right\rangle
%\end{equation}
%
where $\varepsilon=E_1(\varphi_0)-E_2(\varphi_0)$, $\rho_z^0=\rho^0_{11}-\rho^0_{22}\approx \tanh(\varepsilon/2T)$ and $\mathcal{A}=i\mathcal{A}_{12}(\varphi_0)$.
At resonance, $\omega  <\varepsilon_m$, where $\varepsilon_m\sim
\sqrt{D}\Delta_0$ is the MGS bandwidth, we find the dissipative part of the linear
response
\begin{equation}
\gamma_0''(\omega)={\mathrm Im} \gamma_0 = \left(\frac{4e^2N}{C}\right)\omega
g(\omega)\bar{\mathcal{A}}^2\tanh{\omega\over 4T},
\end{equation}
where $\bar{\mathcal{A}}=\mathcal{A}(\varphi_0,\bar{\theta})$, $\bar{\theta}$ being the
resonant angle, $\varepsilon(\bar{\theta})=\omega$. To get a rough estimate of the
dissipation, we  use estimates, $\bar{\mathcal{A}}\sim \kappa\sqrt{\bar D}$, and
$g(\omega)\sim 1/\varepsilon_m$, to obtain,
\begin{equation}\label{MGS}
\gamma_0''(\omega)\sim
\frac{\kappa^2}{R_nC}\left(\frac{\omega}{\varepsilon_m}\right)\tanh{\omega\over4T}.
\end{equation}
It is instructive to compare this result to the dissipative effects of the competing
processes, namely the nodal to nodal state transitions, and the MGS to nodal state
transitions. Below we will show that both the processes produce dissipation, estimated
at zero temperature with
\begin{equation}\label{}
\gamma_{nod}'' \sim \gamma_{nod-MGS}'' \sim
\frac{1}{R_nC}\left(\frac{\omega}{\Delta_0}\right).
\end{equation}
This dissipation is generally small compared to the MGS caused dissipation, by the
factor $\sim \varepsilon_m/\Delta_0$, unless the junction geometry is particularly close
to the symmetric orientation.

%%%%%%%%%%%%%%%
\subsection{Nodal to nodal state transitions }

Applying the general expression, Eq (\ref{linearresponse}), for the linear response to
the nodal quasiparticles we note that the scattering states are four-fold degenerate,
and the indices $i$ label, besides the quasiclassical trajectory angle $\theta$ and the
energy of the incoming quasiparticle $|E|>|\Delta(\theta)|$, also scattering states for
electron/hole-like quasiparticles impinging onto the interface from the left/right
labeled with $s=1,2,3,4$. Remembering that the temporal variation of the phase conserves
the trajectory angle $\theta$, we write contribution of the nodal quasiparticles to the
response at zero temperature on the form,
\begin{equation}\label{gammanod}
\gamma_\text{nod}(\omega)=\frac{4e^2N}{C}\int_{-\pi/2}^{\pi/2} d\theta\int\int
{dEdE'\over |\mathbf{v}_F\cdot
\mathbf{n}|^2}\,\nu(E)\nu(E')\,\frac{\sum_{ss'}\mathcal{A}_{ss'}^2(E,E')\,\varepsilon \,
[\sign(E)-\sign(E')]} {\varepsilon-(\omega+i0)},
\end{equation}
where $\varepsilon=E-E'$, and $\nu(E)$ is the density of states,
\begin{equation}\label{eq.densityofstatesnodes}
\nu(E)= \frac{|E|}{\sqrt{E^2-\Delta^2}}.
\end{equation}

The matrix element for the nodal state transitions is conveniently evaluated through the
relation to the current, similar to the calculation for the MGS,
\begin{equation}
\mathcal{A}_{ss'}(E,E')= \frac{1}{2e}\frac{I_{ss'}(E,E')}{\varepsilon},
\end{equation}
\begin{equation}\label{I2phi} I_{ss'}(E,E')=e\left.\sum_{\sigma}(\mathbf{v}_F\cdot\mathbf{n})
\phi^{\sigma\dag}_s(x,E)\phi^\sigma_{s'}(\tilde{x},E')\right|_{x=0}.
\end{equation}
The sum over scattering states in \Eqs{gammanod}-(\ref{I2phi}) consists of the products
of various scattered waves. Focusing on the most interesting tunnel limit, $D\ll 1$, we
find that the main contribution is coming from the combination of the transmitted and
reflected waves. The current for this combination, $I \sim
e|\mathbf{v}_F\cdot\mathbf{n}|\sqrt{DR}$, is large compared, e.g., to the contribution
of the transmitted waves, $I \sim e|\mathbf{v}_F\cdot\mathbf{n}|\,D$. Therefore
$\mathcal{A}^2\varepsilon \sim |\mathbf{v}_F\cdot\mathbf{n}|^2RD/\varepsilon$, and we
may present the matrix element factor in \Eq{gammanod} on the form,
\begin{equation}
\sum_{ss'}\mathcal{A}^2_{ss'}(E,E')\varepsilon= \frac{D|\mathbf{v}_F\cdot
\mathbf{n}|^2}{\varepsilon} f_{1}\left(\varphi_0,
\frac{E}{\Delta},\frac{E'}{\Delta}\right)+ \mathcal{O}(D^2),
\end{equation}
where $f_1$ is a dimensionless function constructed with the quasiclassical BdG
amplitudes.  With this expression the dissipative part of the response becomes,
\begin{equation}
\gamma_\text{nod}''(\omega)=\frac{4e^2N}{C} \int_{-\pi/2}^{\pi/2} d\theta \,|\Delta|
\int_{1}^{\tilde\omega-1} d\tilde E \ \nu(\tilde E)\nu(\tilde\omega -\tilde E)\,
\frac{D}{\omega}\, f_{1}(\varphi_0,\tilde{E},\tilde\omega - \tilde E).
\end{equation}
where we introduced notations, $\tilde E=E/\Delta$, and $\tilde\omega=\omega/\Delta$. We
see from this equation, that the resonant transitions at $\omega\ll\Delta_0$ select a
small energy interval, $1<\tilde E<\tilde\omega -1$. This imposes a constraint on the
angles, $2|\Delta(\theta)|<\omega$, which are restricted to small areas around the
nodes. This is the source of small value of the dissipation by nodal quasiparticles.

Linearizing the order parameter around this point, $\Delta\approx 2\Delta_0\theta$, and
changing the variable, we finally obtain,
\begin{equation}
\gamma_\text{nod}''(\omega)=\frac{4e^2N}{C}\frac{\omega}{\Delta_0} \int_{2}^{\infty}
{d\tilde\omega\over \tilde\omega^3} \, \int_{1}^{\tilde\omega-1} d\tilde E \ \nu(\tilde
E)\nu(\tilde\omega -\tilde E)\, D\, f_{1}(\varphi_0,\tilde{E},\tilde\omega - \tilde E).
\end{equation}
This integral converges, and gives the estimate for the magnitude of dissipation
produced by the nodal quasiparticles,
\begin{equation}
\gamma_\text{nod}''(\omega)\sim\frac{1}{R_n C}\frac{\omega}{\Delta_0}.
\end{equation}
%

%%%%%%%%%%%%%%%%%%%%%%%%%%%%%%%%%%%%%%%%%%%%%%
\subsection{MGS to nodal state transitions }
The contribution of these processes to the response function at zero temperature  is
given by equation,
\begin{equation}
\gamma_\text{nod-MGS}(\omega)=\frac{4e^2N}{C} \int_{-\pi/2}^{\pi/2} d\theta\int
dE\nu(E)\frac{\sum_{s}\mathcal{A}^2_{s,MGS}(E,E_\text{MGS})
\varepsilon\left[\sign(E)-\sign(E_\text{MGS})\right]}{\varepsilon-\omega-i0}
\end{equation}
where $\varepsilon=E-E_\text{MGS}$, and $E_{MGS}\sim \sqrt{D}\Delta$ is the energy of
the midgap state.
To the lowest order in transparency, the components of the MGS wave functions are
proportional to the factor, $ \sqrt{\Delta/|\mathbf{v}_F\cdot \mathbf{n}|}$, originating
from the wave functions normalization, and they do not depend on transparency.
Evaluating the overlap in the expression for the current matrix elements at the
transmitted side of the scattering state we find that it is proportional to
$e|\mathbf{v}_F\cdot\mathbf{n}|\sqrt{D}\sqrt{\Delta/|\mathbf{v}_F\cdot \mathbf{n}|}$ to
lowest order in transparency.
The transition matrix element can then be written on the form, similar to the nodal to
nodal transitions,
\begin{equation}
\sum_{s}\mathcal{A}^2_{s,MGS}(E,E_\text{MGS})\varepsilon=  D\frac{\Delta
|\mathbf{v}_F\cdot \mathbf{n}|}{\varepsilon}f_{2}\left(\varphi_0,
\frac{E}{\Delta},\frac{E_\text{MGS}}{\Delta}\right)+ \mathcal{O}(D^2),
\end{equation}
where $f_2$ is a dimensionless function of order one.
Using the density of states in Eq. (\ref{eq.densityofstatesnodes})  and the resonance
condition $\varepsilon=E-E_\text{MGS}=\omega$ we present the dissipation on the form,
\begin{equation}
\gamma_\text{nod-MGS}''(\omega)\sim \frac{4e^2N}{C} \int_{-\pi/2}^{\pi/2} d\theta
\,\Theta(E_\text{MGS}+\omega-|\Delta|)\nu\left(\frac{E_\text{MGS}+\omega}{\Delta}\right)D\,
\frac{|\Delta|}{\omega}\,f_{2}\left(\varphi_0,
\frac{E_\text{MGS}+\omega}{\Delta},\frac{E_\text{MGS}}{\Delta}\right).
\end{equation}
We notice that again the resonance condition selects a small angle interval around the
nodes. Furthermore, the MGS energy is small, $E_\text{MGS}/\Delta\sim \sqrt{D}\ll1$, and
can be dropped from the arguments of $\nu$, $\Theta$, and $f_2$. Then we get equation
qualitatively analogous to the dissipation of nodal quasiparticles,
\begin{equation}
\gamma_\text{nod-MGS}''(\omega) =
\frac{4e^2N}{C}\frac{\omega}{\Delta_0}\int_{1}^{\infty}
{d\tilde\omega\over\omega^3}\,\nu(\tilde\omega)\,D\, f_{2}(\varphi_0,\tilde\omega,0),
\end{equation}
or
\begin{equation}
\gamma_\text{nod-MGS}''(\omega)\sim \frac{1}{R_n C}\frac{\omega}{\Delta_0}.
\end{equation}
%

%%%%%%%%%%%%%%%%%%%%%%%%%%%%%%%%%%%%%%%
\section{Nonlinear MGS response}
In this section we  give a derivation of the nonlinear resonant response of the
superconducting phase and MGS to the harmonic temporal oscillation of the current bias,
$I_e(t)=I_e\cos\omega t$. The starting point is the dynamical Eqs. (\ref{phaseeqm}) and
(\ref{densitymatrixeqm}). In what follows, we only focus on the major nonlinear effect
of the MGS transitions, neglecting transitions between the nodal states. The MGS
dynamics is described with a continuum set of two-level density matrices, parameterized
with trajectory angle $\theta$, and satisfying the dynamical Bloch equation,
\begin{equation}
\begin{split}
\dot{\rho}_+&=(-i\varepsilon-\Gamma_2)\rho_++2\dot{\varphi}\mathcal{A}\rho_z\\
%\dot{\rho}_-&=(i\Omega-\Gamma_2)\rho_-+2\dot{\varphi}\mathcal{A}\rho_z\\
\dot{\rho}_z&=-\dot{\varphi}\mathcal{A}(\rho_++\rho_-)-\Gamma_1(\rho_z-\rho_z^0).
\end{split}
\end{equation}
Here the notations are introduced, $\rho_z=\rho_{11}-\rho_{22}$,
$\rho_+=(\rho_-)^*=\rho_{12}$, $\varepsilon=E_1(\varphi)-E_2(\varphi)$, and the
transition matrix element is written on the form
$\mathcal{A}_{12}=i\mathcal{A}(\varphi)$. Phenomenological decay rates
$\Gamma_1,\Gamma_2$ are introduced to account for intrinsic relaxation and dephasing of
the MGS, due to e.g. weak short range disorder.

We consider small oscillations of the phase around the equilibrium value $\varphi_0$, driven by the external current, $I_e(t)=I_e\cos\omega t$, at a frequency not far from the resonant frequency $\delta=\omega-\omega_p\ll 1$.
To separate the slow and fast dynamics we parametrize the phase as:
\begin{equation}\begin{split}
\varphi(t)&=\frac{1}{2}(\varphi_\omega(t) e^{-i\omega t}+c.c.)\\
\dot{\varphi}(t) &=\frac{\omega}{2i}(\varphi_\omega(t) e^{-i\omega t}-c.c.),
\end{split}
\end{equation}
where the complex variable $\varphi_\omega(t)=r(t)e^{i\vartheta(t)}$ depends on the amplitude of oscillations, $r(t)$, and the time dependent phase shift, $\vartheta(t)$.
Using a similar separation for the slow- and fast parts of the off-diagonal elements of
the density matrix,
\begin{equation}
\rho_+(t)=\rho_\omega(t) e^{-i\omega t}, %\qquad \rho_-(t)=\rho_\omega^*(t)e^{i\omega t}
\end{equation}
we get, after expanding to first order in $\varphi-\varphi_0$ and averaging over fast
variables (note $\mathcal{A}_0=\mathcal{A}(\varphi_0)$ and
$\varepsilon_0=\varepsilon(\varphi_0)$),
\begin{equation}\begin{split}
\dot{\rho}_\omega&=-i(\varepsilon_0-\omega-i\Gamma_2)\rho_\omega-i\omega\mathcal{A}_0
\varphi_\omega\rho_z\\
\dot{\rho}_z&=i\frac{\omega}{2}(\varphi_\omega\rho_\omega^*-
\varphi_\omega^*\rho_\omega)-\Gamma_1(\rho_z-\rho_z^0).
\end{split}
\end{equation}
We consider the regime with slow variation of the phase oscillation envelopes,
$\varphi_\omega$, on the the time scale of the MGS decoherence, $\partial_t\varphi \ll
\Gamma_i\varphi$. Then the density matrix will adiabatically follow (in the rotating
frame) the evolution of the phase amplitude, and we consider the quasi-stationary
solutions, $\dot{\rho}_\omega,\dot{\rho}_z\approx 0$,
\begin{equation}
\rho_\omega=%-
\frac{\omega\mathcal{A}_0\varphi_\omega}{\varepsilon_0-\omega-i\Gamma_2}\rho_z,
\end{equation}
\begin{equation}\begin{split}
\rho_z&=%\frac{\left[(\Omega_0-\omega)^2+\Gamma_2^2\right]}{(\Omega_0-\omega)^2+\Gamma_2^2+|\omega\mathcal{A}_0\varphi_\omega|^2(\Gamma_2/\Gamma_1)}\rho_z^0\\
  %&=
\rho_z^0-\frac{(\omega\mathcal{A}_0
r)^2(\Gamma_2/\Gamma_1)}{(\varepsilon_0-\omega)^2+\Gamma_2^2+(\omega\mathcal{A}_0r)^2
(\Gamma_2/\Gamma_1)}\rho_z^0.
  \end{split}
\end{equation}
The nonequilibrium correction to the Josephson current has the form,
\begin{equation}\label{NoneqI}
\text{Tr}\left(\hat{I}_J(\hat{\rho}-\hat{\rho}_0)\right)=2e S\Bigl\langle\partial_\varphi
\varepsilon(\rho_z-\rho_z^0)+\mathcal{A}\varepsilon (\rho^++\rho^-)\Bigr\rangle.
\end{equation}
In the linear approximation with respect to the phase amplitude $r$, only the last term
in \Eq{NoneqI} plays a role, while the corrections to the diagonal matrix elements is of
the second order, $\rho_z-\rho_0\sim \mathcal{O}(r^2)$). Expansion of this term recovers
the result of the linear response calculation Eq. (\ref{mgslinear}),
\begin{equation}\label{}
(2e/C)\text{Tr}\left(\hat{I}_J(\hat{\rho}-\hat{\rho}_0)\right)\approx \omega
\gamma_0(\omega) \varphi_\omega e^{i\omega t}+ c.c.,
\end{equation}
with $\gamma_0(\omega)$  now containing a finite resonance broadening, $\Gamma_2$,
\begin{equation}
\omega\gamma_0(\omega)=\frac{4e^2S}{C}\left\langle
\frac{\omega\mathcal{A}_0^2\varepsilon_0\rho_z^0}{(\varepsilon_0-\omega)-i\Gamma_2}
\right\rangle.
\end{equation}
To go beyond the linear approximation we define, in analogy with the linear response
analysis, a non-linear response coefficient,
$(2e/C)\text{Tr}\left(\hat{I}_J(\hat{\rho}-\hat{\rho}_0)\right)\approx \omega
\gamma(\omega,r) \varphi_\omega e^{i\omega t}+ c.c.$,
\begin{equation}
\omega\gamma(\omega,r)=\frac{4e^2S}{C}
\left\langle\partial_\varphi^2\varepsilon_0(\rho_z(r)-\rho_z^0)+
\frac{\omega\mathcal{A}_0^2\varepsilon_0\rho_z(r)}{(\varepsilon_0-\omega)-
i\Gamma_2}\right\rangle.
\end{equation}
Performing integration assuming the resonance to be narrow, $\Gamma_2,
\mathcal{A}_0\omega r \ll \varepsilon_m$,  we get  for $\gamma=\gamma'+i\gamma''$:
\begin{equation}\begin{split}
\gamma'(\omega,r) &\approx \gamma_0'-
\frac{\partial_\varphi^2\bar{\varepsilon}_0r^2}{\Gamma_1}
\frac{\Gamma\gamma_0''}{\sqrt{(r\bar{\mathcal{A}}_0\omega)^2+\Gamma^2}},\\
\gamma''(\omega,r)&\approx
\frac{\Gamma\gamma_0''}{\sqrt{(r\bar{\mathcal{A}}_0\omega)^2+\Gamma^2}}.
\end{split}
\end{equation}
This is to be inserted into the averaged equation for slow phase amplitude,
\begin{equation}
-i2\omega_p\dot{\varphi}_\omega +\left[-2\omega_p\delta
+\omega_p\gamma(\omega_p,r)\right]\varphi_\omega=\frac{2e}{C}\frac{I_e}{2},
\end{equation}
where $-\omega^2+\omega_p^2\approx -2\delta\omega_p$, or in a more convenient form
\begin{equation}
-i\dot{\varphi}_\omega+\left[-\delta+\frac{\gamma(\omega_p,r)}{2}\right]\varphi_\omega=\frac{\tilde{I}_e}{2},
\end{equation}
where $\tilde{I}_e=(e/C\omega_p)I_e$. Equations for $r(t),\vartheta(t)$ are obtained by
dividing this equation by $\varphi_\omega$ and noticing the relation,
$\dot{\varphi}_\omega/\varphi_\omega=\dot{r}/r+i\dot{\vartheta}$. Identifying real and
imaginary parts yields the set of equations,
\begin{equation}\begin{aligned}
-\frac{\dot{r}}{r}+\frac{\gamma''(r)}{2}&=-\tilde{I}_e\frac{\sin\vartheta}{2r},\\
\dot{\vartheta}-\delta+\frac{\gamma'(r)}{2}&=\tilde{I}_e\frac{\cos\vartheta}{2r}.
\end{aligned}
\end{equation}

The fix points of this set of nonlinear equations is found by solving equation
\begin{equation}
\left[-\delta+\frac{1}{2}\gamma(r)\right]\varphi_\omega=\frac{1}{2}\tilde{I}_e.
\end{equation}
Taking the absolute square of this equation and solving for $\delta$ one finds,
\begin{equation}
\delta=\frac{1}{2}\gamma'(r)\pm \frac{1}{2r}\sqrt{\tilde{I}_e^2-(\gamma''(r))^2 r^2}.
\end{equation}
The two solutions correspond to the stable/unstable branches of the response curve. The
maximum amplitude corresponds to the degenerate point,
$\tilde{I}_e^2=(\tilde{\gamma}'')^2r_m^2$, solution of this equation reads,
\begin{equation}
r_m=\frac{\tilde{I}_e
\Gamma}{\sqrt{\Gamma^2(\gamma''_0)^2-\tilde{I}_e^2\bar{\mathcal{A}}_0^2\omega^2}}\,.
\end{equation}
}

%%%%%%%%%%%%%%%%%%%%%%%%%%%%%%%%%%%%%%

\end{document}